\documentclass[fleqn,usenatbib]{mnras}

\usepackage{newtxtext,newtxmath}
\usepackage{multirow}
\usepackage{threeparttable}

\usepackage[T1]{fontenc}
\usepackage{ae,aecompl}
\usepackage{enumitem}

\usepackage{graphicx}	
\usepackage{amsmath}	
\usepackage{xspace}
\usepackage{ulem}

\newcommand{\angstrom}{\text{\AA{}}}
\newcommand{\rxj}{RX~J1131-1231\xspace}
\newcommand{\he}{HE~0230-2130\xspace}
\newcommand{\q}{Q~2237+0305\xspace}

\title[LSST Microlensing Light Curve Generator]{A Quasar Microlensing Light Curve Generator for LSST}

\author[F. Neira et al.]{
Favio Neira,$^{1}$\thanks{E-mail: fcneirad@gmail.com}
Timo Anguita,$^{1,2}$
and Georgios Vernardos$^{3}$
\\
$^{1}$Departamento de Ciencias Fisicas, Universidad Andres Bello Fernandez Concha 700, Las Condes, Santiago, Chile\\
$^{2}$Millennium Institute of Astrophysics, Chile\\
$^{3}$Kapteyn Astronomical Institute, University of Groningen, PO Box 800, NL-9700AV Groningen, the Netherlands
}

\pubyear{2020}

\begin{document}
\pagerange{\pageref{firstpage}--\pageref{lastpage}}
\maketitle

\begin{abstract}
We present a tool to generate mock quasar microlensing light curves and sample them according to any observing strategy.
An updated treatment of the fixed and random velocity components of observer, lens, and source is used, together with a proper alignment with the external shear defining the magnification map caustic orientation.
Our tool produces quantitative results on high magnification events and caustic crossings, which we use to study three lensed quasars known to display microlensing, viz. \rxj, \he, and \q, as they would be monitored by The Rubin Observatory Legacy Survey of Space and Time (LSST).
We conclude that depending on the location on the sky, the lens and source redshift, and the caustic network density, the microlensing variability may deviate significantly than the expected $\sim$20-year average time scale \citep{mosquera_kochanek2011}.
We estimate that $\sim300$ high magnification events with $\Delta$mag$>1$ mag could potentially be observed by LSST each year.
The duration of the majority of high magnification events is between 10 and 100 days, requiring a very high cadence to capture and resolve them.
Uniform LSST observing strategies perform the best in recovering microlensing high magnification events.
Our web tool can be extended to any instrument and observing strategy, and is freely available as a service at \url{http://gerlumph.swin.edu.au/tools/lsst_generator/}, along with all the related code.
\end{abstract}




\section{Introduction}

Quasars are the most luminous non-transient objects in the sky.
It is currently well understood that they reside in the cores of massive galaxies and that the source of their luminosity is energy released due to matter infalling onto a central supermassive black hole, forming an accretion disc around it that typically shows thermal continuum emission \citep{ss73}.
Due to instabilities in these accretion discs, quasars show intrinsic brightness fluctuations in time scales from hours to years (in the optical and X-ray).
These short variability time scales imply very small accretion disc sizes ($\sim0.001$ pc) that are unresolvable with current instrumentation at typical quasar redshifts (z$\sim$2).
Thus, apart from theory, the precise inner structure of these systems still remains undeciphered.

Quasar microlensing provides us with an exceptional tool to measure the quasar structure at the scale of the accretion disc and the supermasiive black hole environment.
Stars within the lensing galaxies in strongly lensed quasar systems can produce an additional (de)magnification of the multiple "macro"-magnified images.
For a typical system, these stars (or microlenses) have Einstein radii (lensing/magnification scale dependent on mass) that closely match the expected angular sizes of accretion discs at high redshift \citep{chang_refsdal}.
Since the thermal emission of quasar accretion discs is driven by friction of infalling material it has a temperature profile inversely proportional to the radial distance from the central black hole \citep{ss73}.
As such, even though lensing is an achromatic phenomenon, the projected size of accretion disc regions compared to the Einstein radii of the microlenses results in a chromatic signature of microlensing variations: smaller/hotter regions are more effectively (de)magnified than larger/cooler regions.
Analyzing these signatures can thus constrain the physical size and geometry of quasar emission regions \citep[e.g.][]{Irwin1989}.
Indeed, microlensing analysis of about a dozen systems has generally yielded measurements resulting in larger disc sizes and shallower slopes than predicted by standard thin disc theory \citep[e.g.][]{morgan10,jimenez-vicente14,Bate2018}, in agreement with reverberation mapping \citep[e.g.][]{Lira2015,Fausnaugh2016}.

There are two methods to analyze quasar microlensing observations: i) using single epoch broad band or spectroscopic flux ratios \citep[e.g.][]{Bate2008}, or ii) using light curves \citep[e.g.][]{wambsganss1990,webster91,rauch_blandford91,kochanek2004}.
The latter, relies on the fact that the relative velocities between the observer, lens, microlenses, and source produce a change in the alignment and misalignment between the observer, the microlenses and the accretion disc, thus producing (extrinsic) variations of the brightness of the lensed quasar images, in addition to (intrinsic) quasar variability.
The time scales for such brightness fluctuations are from weeks to decades, depending on the specific properties of the system, most importantly the relative velocities and distances.
To constrain the structure of the accretion disc, the most common analysis method consists of comparing observed properties of light curves to those of a statistically representative ($\mathcal{O}\gtrsim 10^6$) sample of simulated light curves for each lensed quasar image \citep[e.g.][]{kochanek2004,anguita08,eigenbrod08,morgan10}.
Long term light curve variations over years and decades allow to constrain simultaneously the dark matter content and mass function of lensing galaxies \citep{kochanek2004,schechter2004,morgan10} in addition to accretion disc structure.
However, the steeper brightness increase (or decrease) shown by single short-term (weeks to months) high magnification events (passing close to, touching, or crossing a microlensing caustic, see Fig. \ref{fig:curve_on_map}) can lead to even more stringent constraints of the accretion disc structure, especially when observed at multiple wavelengths \citep{anguita08,eigenbrod08} due to the strong dependence between source size and magnification profile.

Quasar microlensing studies are not, however, free of limitations.
Firstly, there is a scarcity of strongly lensed quasar systems: there are about $\sim$200 known to date\footnote{\url{https://www.ast.cam.ac.uk/ioa/research/lensedquasars/index.html}}.
Additionally, since monitoring these systems for several years \citep[typical microlensing time scales, see, e.g., ][]{mosquera_kochanek2011} at high cadence (to map the slope of magnification events) with accurately resolved photometry is not an easy task. Monitoring data is only available for a handful of systems and mostly in single photometric bands, e.g. the COSMOGRAIL program\footnote{\url{https://www.epfl.ch/labs/lastro/scientific-activities/cosmograil/}} \citep{COSMOGRAIL2020}.
Even when such data is available, statistically comparing these observed light curves to the huge (and degenerate) parameter space covered by microlensing simulations is a computationally exhaustive task usually tailored to specific individual systems \citep{kochanek2004,Poindexter2010}.
However, the forthcoming large area surveys in the next decade, like The Rubin Observatory Legacy Survey of Space and Time (LSST), are expected to increase the number of known lensed quasars by at least one order of magnitude \citep{OM10} and provide multiband monitoring for thousands of systems.
This certainly will tackle the lack of microlensing data, but will naturally increase the computational cost of the analysis.
As such, it is becoming necessary to optimize the analysis or select sub-samples of those newly discovered systems that are likely to provide the most stringent constraints on the inner structure of accretion discs.

In this work, we present a tool to efficiently simulate microlensing light curves in several photometric bands for any lensed quasar, while properly taking into account all the relevant parameters, like accretion disc models, the relative velocities, distances, and lens mass model parameters.
Our simulations can also take into account several observational constraints and cadences specific to monitoring campaigns and/or surveys.
We specifically examine realistic LSST-like light curves for Gaussian shaped accretion discs, however, our tool is versatile enough to include new developments in accretion disc models and monitoring campaigns.
In Section 2, we introduce the LSST and the several possible observing strategies that it may follow.
In Section 3, we describe the ``LSST Microlensing Light Curve Generator'' along with all the relevant physical parameters and the way we generate mock light curves.
Section 4 presents a test case where we generate 10000 light curves for 3 well-studied yet diverse systems that have shown evidence of microlensing, calculating their expected microlensing properties and benchmarking the different candidate LSST observing strategies in their ability to sample high magnification events.
Finally, in Section 5, we discuss the capabilities and limitations of our generator.
In this work we considered a flat $\Lambda$CDM cosmology with $H_0 = 67.7$ km s$^{-1}$ Mpc$^{-1}$, $\Omega_m = 0.309$ and $\Omega_\Lambda = 0.691$.

\section{The LSST Observing Strategies}
\label{sec:lsst}
The LSST is a ground-based optical telescope currently under construction at Cerro Pach\'on in Northern Chile.
It will have an 8.4-m primary mirror with a field of view of 9.2 deg$^2$, which will allow it to repeatedly observe the southern sky continuously over 10 years in 6 photometric bands (u, g, r, i, z, y).
Thus, given the rarity of lensed quasars, the expected quasar microlensing time scales, and the chromatic nature of the events, the LSST is potentially the ideal quasar microlensing probe of the next decade.

However, it is important to note that the specific observing strategy that LSST will follow is, at the time of writing, not completely defined and might not be sufficient to fully probe all the quasar microlensing variability of interest, e.g. high magnification events.
Even though LSST has a broad spectrum of science goals, from solar system science to cosmology, there are just four core science goals\footnote{LSST's core goals are: i) constraining dark energy and dark matter; ii) making an inventory of the Solar System; iii) exploring the transient optical sky; and iv) mapping the Milky Way.} that yield the following overarching quantitative constraints on the observing strategy  \citep{LSSTObservingStrategyWhitePaper}:
\begin{itemize}[leftmargin=*]
    \item A footprint for the ``main survey'' of at least 18000 deg$^2$, which must be uniformly covered to a median of 825 30-s long visits per 9.6 deg$^2$ field, summed over all six filters. This places a minimum constraint on the time required to complete the main survey, with simulations indicating a required 85 to 90 per cent of the available time (10 years) to reach this benchmark.
    \item Parallax and proper motion 1$\sigma$ accuracy of 3 milli-arcsec and 1 milli-arcsec per year per coordinate at r = 24, respectively, in the main survey. This places a weak constraint on how visits are distributed throughout the lifetime of the survey and throughout a season.
    \item Rapid field revisits (40 s to 30 min) must be acquired over an area of at least 2000 deg$^2$ for very fast transient discovery. This requirement can usually be satisfied via simple field overlaps when surveying contiguous areas of sky.
\end{itemize}

These ``main'' constraints can be achieved with several different observing strategies.
The LSST simulations team, through their LSST operations simulator (OpSim), has provided the community with several simulated observing strategies that, even when achieving these high level constraints, can have significant consequences for other science cases (like quasar microlensing).
The LSST project has encouraged the community to test different observing strategies against additional science cases in order to converge into a strategy that maximizes the scientific output.

Disregarding the subtle differences among the different observing strategies, they can be classified into two main families:
\begin{itemize}[leftmargin=*]
    \item Uniform cadences: Intended to maximize uniformity in both cadence and filters within the high level constraints. As such, this cadence family would maximize the mapping of longer time scale brightness fluctuation events that can be seen in all LSST bands.
    \item ``Rolling'' cadences: Intended to maximize shorter time scale events. This is generally achieved by focusing on specific areas of the sky (and sometimes observing bands) for a significant period of the survey and then focusing on (or ``rolling to'') a different area (and/or band).
\end{itemize}
Despite the existing ones, any observing strategy can be fed into our light curve generator to obtain realistically simulated light curves, as described in the following section.

\section{Light Curve Generator}
\label{sec:generator}
Much like in most quasar microlensing quantitative studies, to generate light curves that are representative of a particular lensing system one needs to: assume a macromodel for the lens, create corresponding microlensing magnification maps, assume a light profile and spectral distribution for the source, and a model for the velocities involved \citep[e.g.][]{kochanek2004,anguita08,eigenbrod08}.
This leads to continuous ``theoretically expected'' light curves, which, in order to be converted to mock observations, need to be sampled according to an instrument's specifications and observing strategy.
Here we describe our treatment of these forward modelling steps to generate generic theoretical light curves and then sample them according to some observational strategy.

\subsection{Magnification maps}
A macromodel of the lens determines the global mass distribution based on the relative positions of the lensing galaxy and the quasar multiple images (occasionally on their magnifications as well), and/or the quasar host galaxy's extended lensed features.
This model provides three main parameters as a function of position in the sky with respect to the mass distribution that are used to define the local microlensing properties: the convergence, $\kappa$, the shear, $\gamma$, and the smooth matter fraction, $s$. These respectively define the total surface mass density, the shearing applied to the deflection angles due to tidal effects, and the fraction of the mass density into compact (matter in form of stellar-mass objects) and smooth (dark) matter.

Based on macromodel estimates of the $\kappa,\gamma,$ and $s$, the collective effect of the deflection angles of an ensemble of microlenses within the main lensing galaxy can be calculated.
This is a computationally demanding task, for which several variants of the inverse ray-shooting technique \citep{Kayser1986} have been proposed, benefiting from optimized software \citep{Wambsganss1999,Mediavilla2011a} and hardware solutions \citep{Thompson2010}.
The end result is a grid on the source plane that describes the magnification per pixel due to the microlensing effect - the magnification map.
For most applications, and if the magnification maps are wide enough, a single random realization of the microlens positions per $\kappa,\gamma,s$ is sufficient \citep{Vernardos2013}.

The scaling of the deflection angles depends purely on the Einstein radius, which is a measure of the lensing power of the microlenses, given by:
\begin{equation}
\label{eq:rein}
R_{\rm E} = \sqrt{ \frac{D_{\rm os}D_{\rm ls}}{D_{\rm ol}} \frac{4G\langle M \rangle}{c^2} } \, ,
\end{equation}
where $D_{\rm ol}$, $D_{\rm os}$, and $D_{\rm ls}$, are the angular diameter distances from observer to lens, observer to source, and lens to source respectively, $\langle M \rangle$ is the mean mass of the point-mass microlenses, $G$ is the gravitational constant, and $c$ is the speed of light.

The inverse ray shooting process is a computationally very expensive task, therefore, we take advantage of the pre-computed magnification maps from the GERLUMPH\footnote{\url{http://gerlumph.swin.edu.au}} resource \citep{vernardos2014} that consists of $\sim$75000 maps over a wide range of $\kappa,\gamma,s$.
The maps are 25 $R_{\rm E}$ and 10000 pixels on each side, resulting in a resolution of 0.025 $R_{\rm E}/$pixel, and have been computed with a fixed mass for all the microlenses; selecting a specific mass value sets the length dimension of the maps through equation (\ref{eq:rein}). Although distributing the baryonic mass to an ensemble of stellar mass objects using a given Initial Mass Function prescription is physically justified, it has been shown to have a minimal effect \citep{wambsganss1990,lewis1995,wyithe2001}.

\subsection{Accretion Disc Profile}
In order for microlensing to have any noticeable effect, the light emitting region of the source has to be comparable in angular size with the corresponding $R_{\rm E}$ of the microlenses, i.e. $\sim \mu$arcsec.
For quasar sources, this corresponds to the region extending from the immediate neighbourhood of the central supermassive black hole to the broad-emission line region, and from X-rays to the infrared part of the spectrum (taking redshift into account).
Here we focus on the accretion disc and its emission in the optical and UV.

An important theoretical quantity that can be observed in the case of microlensing is the temperature profile of the accretion disc.
The general picture is that there is a radial temperature gradient, with the innermost parts being the hottest, and the disc being locally in thermal equilibrium.
The thin disc model \citep{ss73} is the standard theoretical description of such a system:
\begin{equation}
\label{eq:ss-disc}
r_{1/2} = 9.7 \times 10^{15} \left( \frac{M_{\rm BH}}{10^{9} M_{\odot}} \right)^{2/3} \left( \frac{f_{\rm Edd}}{\eta} \right)^{1/3} \left( \frac{\lambda}{\mu m} \right)^{4/3} \quad \mathrm{cm},
\end{equation}
where $r_{1/2}$ is the radius containing half the luminosity, i.e. the half-light radius, at the rest wavelength $\lambda=\lambda_{\rm obs}/(1+z_s)$ (we have replaced temperature with wavelength due to the thermal equilibrium assumption), $\lambda_{\rm obs}$ is the observed wavelength, $z_s$ is the source redshift, $M$ is the mass of the supermassive black hole, $f_{\mathrm{Edd}}$ is the Eddington luminosity, and $\eta$ the accretion efficiency.
However, observations seem to differ from this basic theoretical model, finding both steeper and shallower temperature profiles \citep[e.g.][]{Bate2008,Floyd2009,Blackburne2011,Rojas2014}.
A more flexible model with respect to this, is the following power law:
\begin{equation}
\label{eq:rhalf}
r_{1/2} = r_0\left(\frac{\lambda}{\lambda_0}\right)^\nu,
\end{equation}
where $r_0$ is the half-light radius at the reference wavelength $\lambda_0$, and $\nu$ is the power law index that determines the size dependence on $\lambda$ \citep[see also][]{jimenez-vicente14,Bate2018}.

Equation (\ref{eq:rhalf}) is a measure of the mean size in a given wavelength, but the actual light distribution can vary in shape, from being a smooth or even flat function of radius to a detailed profile having features like gaps, wraps, hotspots, etc.
In the following we model the disc in each wavelength as a two-dimensional Gaussian profile.
It turns out that the actual shape of the profile plays a small role in the average microlensing properties \citep{Mortonson2005,Vernardos2019}.

Given $z_s$ and some $r_0$, $\lambda_0$, $\nu$, or $M_{\rm BH}$, $f_{Edd}$, $\eta$, we use equation (\ref{eq:rhalf}) or (\ref{eq:ss-disc}) to create the profile of the source in any wavelength.
Once the spatial brightness profile of the accretion disc is compared to the magnification map, it is very likely that it will extend beyond a single pixel, in which case the map needs to be convolved with the profile.
If instead of an analytic a numerical profile of the accretion disc is used, then creating a two-dimensional projection of that profile in a grid matching the resolution of the magnification map is required (which is also done for the analytic profiles in order to perform the convolutions).
Obviously, in both cases the value of $R_{\rm E}$ needs to be fixed to set the map physical dimensions.

\subsection{Velocity model}
The time-varying microlensing signal is due to varying magnification of the source assuming that it moves across the magnification map.
The effective velocity of the source, $\boldsymbol{\upsilon}_e$, is thus the vectorial sum of the transverse (on the sky plane) velocities of the observer, $\boldsymbol{\upsilon}_o$, microlenses, $\boldsymbol{\upsilon}_l$, and source, $\boldsymbol{\upsilon}_s$, measured in the observer's rest frame and projected on the source plane \citep{Kayser1986}.
In principle, the caustic pattern itself can move due to the proper motions of the microlenses leading to ``faster'' variability \citep[e.g.][]{Kundic1993}, however, \citet{Wyithe2000} have shown that this effect can be approximated by a bulk velocity for all the microlenses based on the lens velocity dispersion, whose magnitude is:
\begin{equation}
    \label{eq:vstar}
    \upsilon_\star = \sqrt{2} \epsilon \sigma_\star,
\end{equation}
where $\sigma_\star$ is the velocity dispersion at the lens center and $\epsilon$ is an efficiency factor depending on $\kappa,\gamma$ \citep[see ][for details, here we assume $\epsilon=1$]{Wyithe2000}.
The direction of $\boldsymbol{\upsilon}_\star$ is random.

The microlens velocity component, $\boldsymbol{\upsilon}_l$, is the vectorial sum of $\boldsymbol{\upsilon}_\star$ and the peculiar velocity of the lens galaxy.
The latter has a random direction and a magnitude drawn from a normal distribution of specified width.
This is also true for the peculiar velocity of the source - the only source velocity component.
Therefore, we can combine these two normal random variables to a single velocity component, which is random in direction and has a magnitude drawn from a Gaussian distribution with standard deviation given by:
\begin{equation}
    \label{eq:vgauss}
    \sigma_g = \left[ \left( \frac{\sigma_l^{\mathrm{pec}}}{1+z_l}\frac{D_{os}}{D_{ol}} \right)^2 + \left( \frac{\sigma_s^{\mathrm{pec}}}{1+z_s}\right)^2 \right]^{1/2},
\end{equation}
where $\sigma_l^{\mathrm{pec}}$ and $\sigma_s^{\mathrm{pec}}$ are the standard deviations of the peculiar velocity distributions of the lens and the source respectively.

\begin{figure}
	\includegraphics[width=\columnwidth]{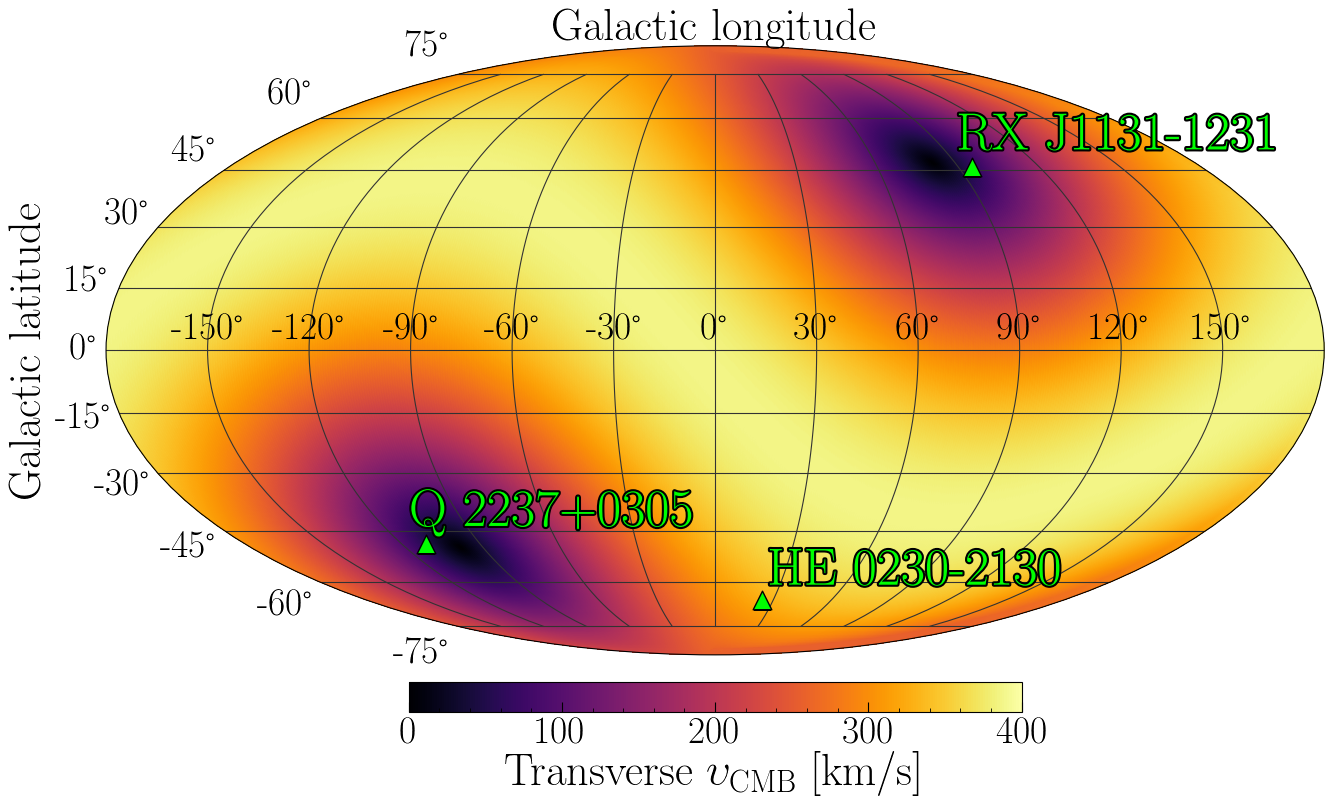}
	\caption{Magnitude of the transverse velocity of the observer, $\boldsymbol{\upsilon}_{0}$ (equation \ref{eq:vcmb}), which is assumed the same as the velocity of the CMB, as a function of sky coordinates. The locations of the three lensed quasars examined in this work are also indicated.}
	\label{fig:obsvel}
\end{figure}

The transverse velocity of the observer is defined with respect to the Cosmic Microwave Background (CMB) velocity dipole \citep{Kogut1993}:
\begin{equation}
    \label{eq:vcmb}
    \boldsymbol{\upsilon}_0 = \boldsymbol{\upsilon}_\mathrm{CMB} - (\boldsymbol{\upsilon}_\mathrm{CMB} \cdot \hat{z})\hat{z},
\end{equation}
where $\boldsymbol{\upsilon}_\mathrm{CMB}$ is the CMB velocity - a known function of sky coordinates - and $\hat{z}$ represents the observer's line of sight.
As such, the velocity of the observer is exactly defined both in magnitude and direction by the coordinates of the lens system (see Fig. \ref{fig:obsvel}).

Combining the explicit velocity components we finally get:
\begin{equation}
    \label{eq:veff_explicit}
    \boldsymbol{\upsilon}_e = \frac{\boldsymbol{\upsilon}_o}{1+z_l}\frac{D_{ls}}{D_{ol}} - \frac{\boldsymbol{\upsilon}_\star}{1+z_l}\frac{D_{os}}{D_{ol}} + \boldsymbol{\upsilon}_g,
\end{equation}
where $\boldsymbol{\upsilon}_o$ has a fixed magnitude and direction given by equation (\ref{eq:vcmb}), $\boldsymbol{\upsilon}_\star$ has a random direction and magnitude given by equation (\ref{eq:vstar}), and $\boldsymbol{\upsilon}_g$ has a random direction and magnitude sampled from a Gaussian with width given by equation (\ref{eq:vgauss}).
The velocity model is thus completely defined by determining the sky coordinates of the system (and hence $\boldsymbol{\upsilon}_{\mathrm{CMB}}$), $\sigma_\star$, $\epsilon$, $\sigma_l^{\mathrm{pec}}$, and $\sigma_s^{\mathrm{pec}}$.

The final step in determineing the velocity vector with respect to the magnification map, is finding the correct orientation of the map on the plane of the sky.
This is given by the angle of the shear, $\phi_\gamma$, at the corresponding location on the lens plane, i.e. where a multiple quasar image lies \citep[e.g. see figure 1 of][]{Poindexter2010}.
A crude approximation for this angle can be the direction to the center of brightness of the lens galaxy, which is exact if there is no external shear and the center of light coincides with the center of mass.

\subsection{Producing mock light curves}
With the components needed to generate light curves in place, we now discuss how we combine them and include the instrumental/observing strategy effects.
We note that the methodology described here is not LSST-specific and can be extended to any instrument/monitoring campaign.

First, we need to have a profile of the source in each of the 6 photometric LSST bands: u, g, r, i, z and y ($\lambda_{eff}$= 3654.9, 4800.3, 6222, 7540.6, 8682.1 and 9925 \angstrom).
We create these by matching the band wavelengths converted to the quasar rest-frame to the $r_{1/2}$ for the source, either from equation (\ref{eq:rhalf}) or equation (\ref{eq:ss-disc}), and selecting a shape for the disc profile, viz. a two-dimensional Gaussian or a uniform disc, or using custom profiles.
By convolving the profiles with the magnification map, we end up with 6 convolved maps from which the light curves will be extracted.

To combine the convolved maps with the velocities, a random sample of $\boldsymbol{\upsilon}_{e}$ is generated from the velocity model (equation \ref{eq:veff_explicit}) together with a corresponding set of random starting points on a map.
This completely defines a set of trajectories on a map.
The maximum length of the light curves that can be extracted from each trajectory is determined by the time spanned by the observations, and varies because the magnitude of $\boldsymbol{\upsilon}_{e}$ is not fixed.
Magnification values from the convolved maps can be extracted at any given time step by finding the corresponding location along a trajectory and doing a bilinear interpolation on the map pixels.
In this way, two kinds of light curves can be created: full resolution light curves, as they would be observed by continuous monitoring (limited only by the map resolution), and light curves sampled at irregular intervals corresponding to some observing strategy (affected by season gaps, clouded skies, etc.).

\begin{figure*}
\includegraphics[width=\textwidth]{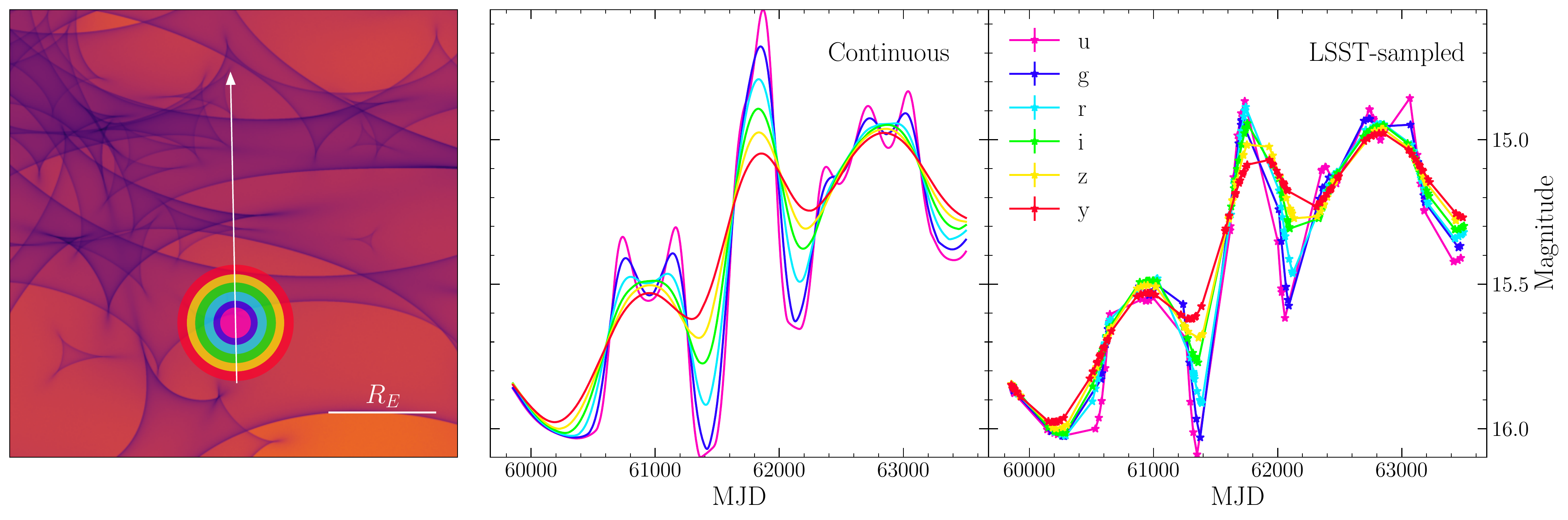}
\caption{Light curves for \q, generated using the parameters listed in Table \ref{tab:parameters} and sampled according to a given observing strategy (colossus\_2664).}
\label{fig:curve_on_map}
\end{figure*}

In order to produce continuous light curves, we set a time interval equal to the width of a pixel and extract magnification values from the beginning to the end of a trajectory (set by some $t_{\rm max}$ - 10 years in the case of LSST).
These time steps are the same for all observed wavelengths, i.e. for all the convolved maps.
The time resolution of the light curves obviously depends on the resolution of the magnification map, which can vary among different lens systems; for all cases examined here the map resolution is sufficiently high, however, higher resolution may be required when examining detailed profiles of an accretion disc crossing a caustic, e.g. in X-rays.
An example of continuous light curves is shown in the middle panel of Fig. \ref{fig:curve_on_map}.

As described in Section \ref{sec:lsst}, the LSST observations will neither happen in regular intervals nor at the same time in all the bands: each pointing of the telescope at a specific area of the sky where a lens system happens to lie is subject to the observing strategy resulting from a number of factors and constraints in a complicated way.
Using LSST OpSim output observing strategies, we can determine the times of the observations in each band and sky coordinate, create a corresponding time sequence, and use it to extract realistically sampled light curves.
An example of such light curves is shown in the right panel of Fig. \ref{fig:curve_on_map}.
Finally, for each trajectory there will be one continuous and one sampled light curve associated with each band - a total of 12 light curves.

The magnification values in the maps are, by construction, relative to the macro-magnification, and, therefore, they will need to be scaled accordingly.
However, the true magnitude of the source is unknown, and the macro-magnification is known to have a large uncertainty due to degeneracies in the macromodel, or the presence of massive substructures in the lens that can modify it without affecting microlensing.
In addition, absorption can happen in the lens creating a varying magnitude of the source across wavelengths.
For these reasons, in order to bring the light curve magnitudes to scale the true (unmicrolensed) magnitude of the source in each band has to be given.
This will also be used to estimate the errors of the LSST photometry in each band  and is based purely on instrumental/site parameters \citep[see eq. 5 in][]{LSST2}.

\subsection{The online tool}
We created a user-friendly web interface for the light curve generator, accessible at:
\begin{center}
    \url{http://gerlumph.swin.edu.au/tools/lsst_generator/},
\end{center}
that can be used to send requests for LSST simulated light curves to the GERLUMPH infrastructure.
The tool automatically performs all the resource-heavy computations, collects and packages the output, and sends a notification and download link when the requested data are ready.

The input required by the user consists of the parameters of the theoretical model components described in this section, i.e. the magnification map, the source, and the velocity model, together with those related to the LSST observing strategy.
Regarding the source, apart from the analytic profiles described above there is the possibility to upload user-generated accretion disc profiles.

The output of the code is a statistical sample of light curves that represent the system under examination.
Along with the total observing time, the user needs to specify the desired number of trajectories.
The basic unit of output is a light curve object, defined simply as a list of magnitude, mag, and magnitude uncertainty, $\delta$mag (only for sampled light curves), values as a function of time, $t$.
The two types of light curves, continuous and sampled, differ only in the sequence of time steps: the former have regular time intervals, while the latter have a sequence of time steps distinct for each band but the same for all trajectories.
Further detailed description of the output can be found at the webpage of the tool.

\subsubsection{Implementation}
The tool consists of three parts that work together: the web-based user interface used to set the input parameters, a high-performance code that performs all the simulations, and a fully automated `supervisor' program invisible to the user, which enables the communication and exchange of data.
The simulations code is further divided into a part that calculates the theoretical model, which is written in C++ using the \textit{gerlumph++}\footnote{\url{https://github.com/gvernard/gerlumphpp}} library and allows for additional speed gains from Graphics Processing Unit hardware acceleration to perform the convolutions, and a part using the Python-based LSST Simulation Framework\footnote{\url{https://confluence.lsstcorp.org/display/SIM/}} that provides the observing strategy.
Both codes are provided as open-source software, in order to enable the reproducibility of our scientific results\footnote{\url{https://github.com/gvernard/lsst_generator}}.

The automated pipeline is embedded within the GERLUMPH infrastructure/framework \citep{Vernardos2014b}.
This includes software - a database, database management and interface programs, additional visualization tools, and a user notification system - and hardware provided by Swinburne University, consisting of a web server and the OzSTAR supercomputer\footnote{\url{https://supercomputing.swin.edu.au/}}, whose GPUs are the real workhorse behind the tool.
The entire software pipeline is portable and does not depend explicitly on the hardware: the main pieces of software are freely available and can be deployed locally at the users' end as a stand-alone version of the tool.

\section{Trial application}
To test our light curve generator, we selected three systems that are known for showing evidence of microlensing: \q \citep{Irwin1989}, \he \citep{wisotzki99} and \rxj \citep{sluse2006}.
These systems were selected to sample a broad range of relative distances between observer, lens, and source, as well as sky coordinates (see Fig. \ref{fig:obsvel}), all of which have direct influence in the resulting transverse velocities of each light curve track.
For each system we investigated only one image (image A) and produced $10000$ 10-year-long light curves in the u, g, r, i, z and y LSST filters.

\begin{figure*}
\includegraphics[width=\linewidth,height=\textheight,keepaspectratio]{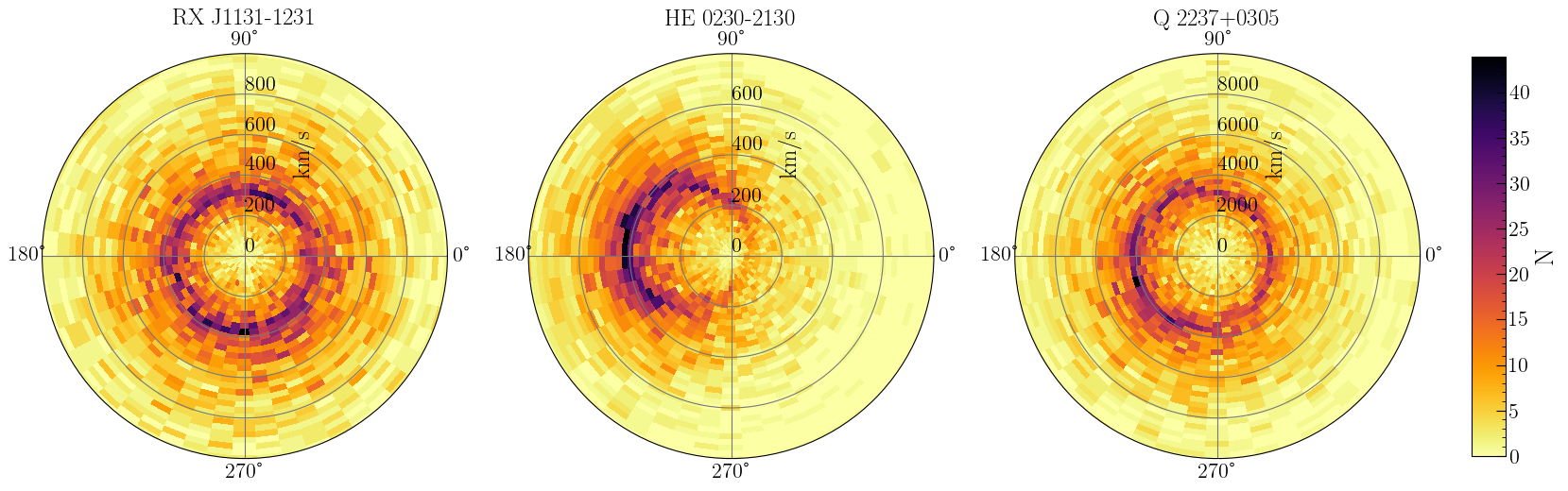}
\caption{Two-dimensional histogram (number counts N) of the effective velocity magnitude and direction for 10000 light curves generated for each system. The mean velocity magnitudes are $488$, $384$ and $4628$ km/s for \rxj, \he and \q respectively.}
\label{fig:angvelsim}
\end{figure*}

\begin{figure*}
\includegraphics[width=\textwidth,keepaspectratio]{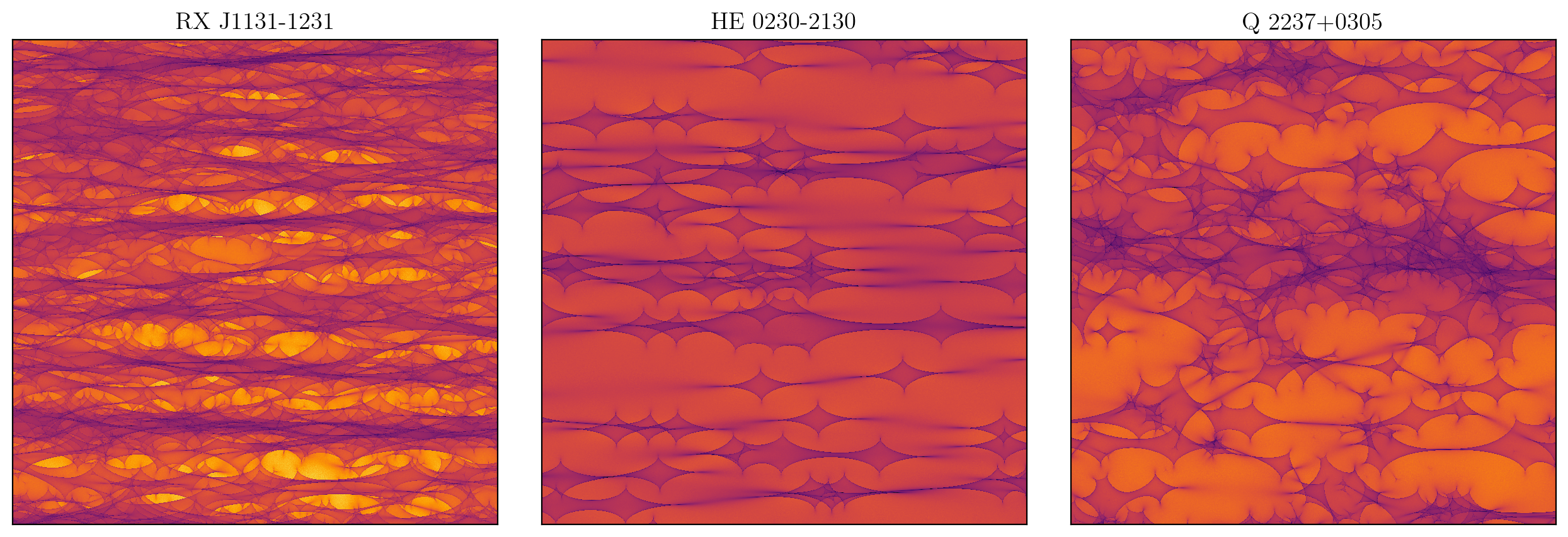}
\caption{Magnification maps for the 3 systems. The map for \rxj corresponds to a minimum image and the other two to saddle-points, which, together with the smooth matter fraction, define the caustic density on the maps (see the parameters listed in Table \ref{tab:parameters}).}
\label{fig:maps}
\end{figure*}

The scaling of accretion disc size with wavelength was fixed to a thermal profile with slope $\nu$=4/3, assumedidentical for all three systems for simplicity. To be able to compare our results with previous estimates, we have adopted the size $r_0$ of the accretion disc from the theoretical estimates of \citet{mosquera_kochanek2011} at restframe i-band wavelength (assumed face-on).
The peculiar velocities of the lensing galaxies and background quasars were drawn from a normal distribution with $\sigma_l^{\mathrm{pec}}=\sigma_s^{\mathrm{pec}}=250$ km/s at their respective planes, while the velocity dispersion of the lensing galaxies was fixed to $200$ km/s.
The resulting transverse velocities of the $10000$ tracks for each system are shown in Fig. \ref{fig:angvelsim}.
The remaining parameters for the systems, i.e. sky coordinates, redshifts, and macroscopic parameters ($\kappa,\gamma$) were obtained from the literature \citep{poindexter_kochanek,schechter2014}, while the smooth matter fraction, $s$, was chosen to be consistent with the work of \cite{pooley2012}.
The corresponding magnification maps are shown in Fig. \ref{fig:maps}.
All the parameters that define the simulated light curves are listed in Table \ref{tab:parameters}.

\begin{table}
\centering
\caption{Parameters required to generate mock light curves, described in Section 3.}
\label{tab:parameters}
\begin{tabular}{|l|l|l|l|}
\hline
                                   & {\rxj}  & {\he}     & {\q}     \\
\hline
$r_0\text{[ld]}$                   & 0.01             & 0.01               & 0.07              \\ 
$\lambda_0\text{[nm]}$             & 102.6            & 102.6              & 102.6             \\
$\nu$                              & 4/3              & 4/3                & 4/3               \\
$\kappa$                           & 0.49             & 0.47               & 0.39              \\ 
$\gamma$                           & 0.56             & 0.42               & 0.40              \\ 
s                                  & 0.60             & 0.90               & 0.00              \\ 
$\langle$M$\rangle\text{[M}_\odot]$ &0.3&0.3&0.3\\
$\text{z}_\text{s}$                & 0.658            & 2.162              & 1.695             \\ 
$\text{z}_\text{l}$                & 0.295            & 0.52               & 0.034             \\
$\sigma_\star\text{[km/s]}$        & 200              & 200                & 200               \\
$\sigma_l^{\mathrm{pec}}\text{[km/s]}$   & 250              & 250                & 250               \\
$\sigma_s^{\mathrm{pec}}\text{[km/s]}$   & 250              & 250                & 250               \\
RA [hours]                                 & 11.50           & 2.54              & 22.68            \\ 
DEC [deg]                         & -12.50           & -21.29        	   & 0.0$^\star$              \\ 
$\phi_\gamma\text{[deg]}$          & 106.08           &  -176.20            & 175.43             \\
\hline
\end{tabular}
\medskip

$^\star$We shift the declination of \q from its true value at 3.36 to 0.0 in order to avoid the fact that certain LSST observing strategies exclude the northern part of the sky. This small offset does not affect our results.

\end{table}

\begin{figure}
 \includegraphics[width=\linewidth,height=\textheight,keepaspectratio]{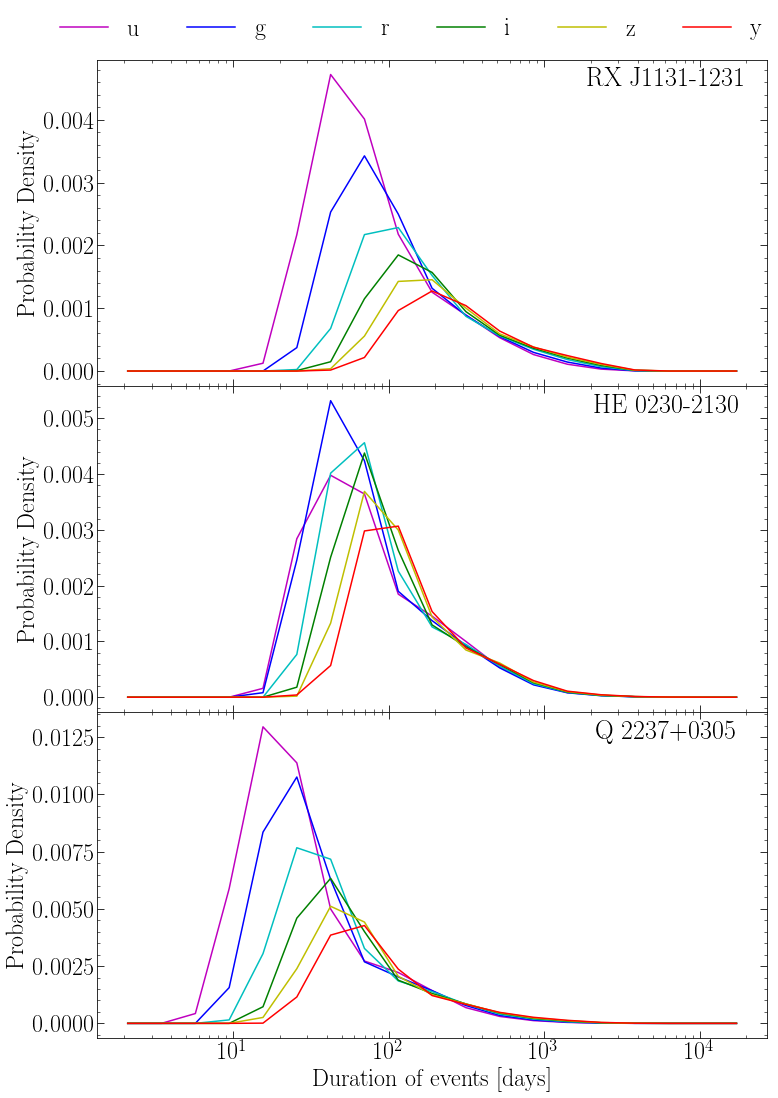}
\caption{Probability density of the duration of high magnification events ($\Delta$mag $>1$ in the r band), defined as the time span between the peak of the event and 10 per cent of its value.}
\label{fig:eventdays}
\end{figure}

We measured the number of high magnification events in the continuous light curves, where we define a high magnification event as any brightening greater than $\Delta$mag=1.
Table \ref{tab:nevents} shows the average number of high magnification events in each of the simulated bands for each system per decade.
Additionally, we also measured the duration of the events as shown in Fig. \ref{fig:eventdays}, defined as the time span between the magnification at the peak of the event and 10 per cent of its value.

To further investigate the nature of the high magnification events, we identified the number of caustic crossings (center of the track intersecting a caustic) expected in 10 years for each system (see Table \ref{tab:nevents}).
As in \citet{anguita08}, this was performed using the analytical method by \citet{witt90} to precisely identify the location of each caustic in the magnification map.

\begin{figure}
\includegraphics[width=\linewidth,height=\textheight,keepaspectratio]{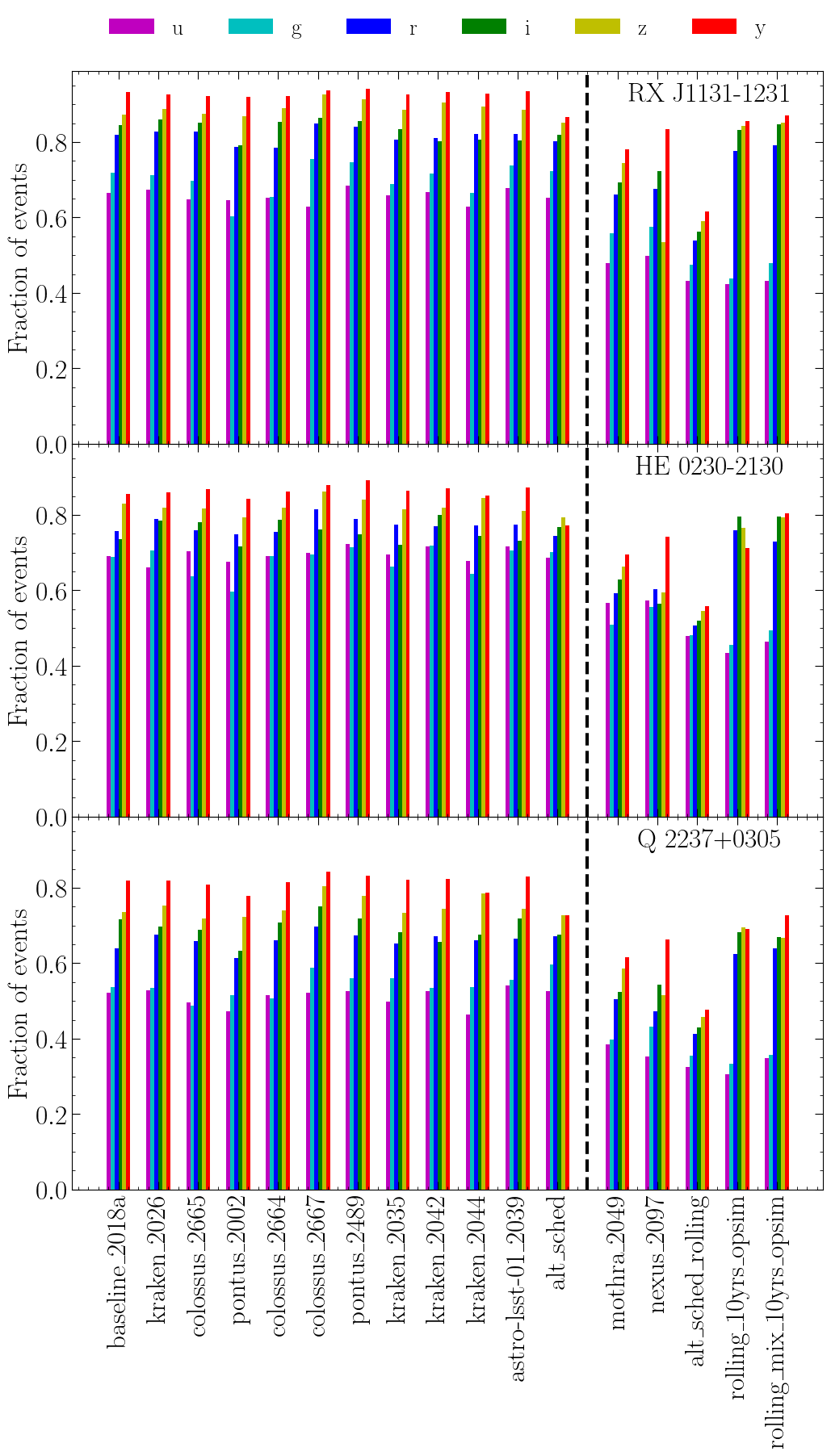}
\caption{Detection performance of high magnification events ($\Delta$mag $>1$ in the r band) in 10 years of LSST monitoring. The vertical dashed line splits ``uniform'' (left) and ``rolling'' (right) cadences.}
\label{fig:opsimplots}
\end{figure}

In order to test the relative performance of the different LSST observing strategies in recovering the expected high magnification events, we independently counted the events in the LSST-sampled versions of the continuous light curves.
We have specifically selected the OpSim runs prepared for the Call for White Papers on LSST Cadence Optimization\footnote{\url{http://ls.st/doc-28382}}, in addition to outputs from the popular alternative scheduler alt\_sched\footnote{\url{http://altsched.rothchild.me:8080/}} as well as two ``rolling cadence'' experiments provided by the LSST simulations team\footnote{\url{https://community.lsst.org/t/replacement-rolling-cadence-simulations/3321}} shortly after the Call\footnote{\url{https://www.lsst.org/call-whitepaper-2018}}. The sample thus comprises a total of 12 uniform and 5 rolling cadences.
Fig. \ref{fig:opsimplots} shows the ratio between the number of magnification events detected in the LSST-like observed light curves and in the continuous light curves for each band and observing strategy studied. Note that we only take into account the brightness fluctuations lost due to temporal gaps resulting from the observing strategy, as the photometric uncertainties are completely negligible for events with brightness variations larger than 1 mag.

We note that this experiment has not considered uncertainties in any of the input parameters (Table \ref{tab:parameters}).  As such, the uncertainties shown in Table \ref{tab:nevents} are exclusively due to the variance of the statistically representative sample of light curves. Even though some of the input parameters can be more measured through more accurate lens models using high resolution imaging and spectroscopy, the microlens mass function at typical lens redshifts (s and $\langle$M$\rangle\text{[M}_\odot]$) and the accretion disc structure at typical source redshifts ($r_0$ and $\nu$) are not. As discussed in the following section, these certainly influence the total number and properties of caustic crossings and high magnification events, respectively \citep[see also][]{pooley19,moustakas19}.

\begin{table}
\caption{Average number of high magnification events ($\Delta \mathrm{mag}>1.0$) in each of the LSST bands for 10 years of monitoring time for each system. The expected number of caustic crossings is also listed. The limits indicate the $1\sigma$ interval of the number of high magnification events. For the cases where having 0 events/caustic crossings is consistent within 1$\sigma$ we have dropped the lower limit.}

\def\arraystretch{1.5}
\label{tab:nevents}
\centering
\begin{tabular}{|l|l|l|l|}
\hline
                    & {\rxj}                      & {\he}                       & {\q}                        \\
\hline
u                   & 1.890$^{+0.982}_{-1.690}$   & 0.495$^{+0.578}$   & 2.332$^{+1.605}_{-1.893}$   \\ 
g                   & 1.599$^{+0.689}$            & 0.488$^{+0.587}$   & 2.105$^{+0.963}$            \\ 
r                   & 1.328$^{+0.660}$            & 0.467$^{+0.616}$   & 1.859$^{+0.560}$            \\ 
i                   & 1.171$^{+0.649}$            & 0.444$^{+0.643}$   & 1.662$^{+0.580}$            \\ 
z                   & 1.048$^{+0.624}$            & 0.421$^{+0.676}$   & 1.524$^{+0.544}$            \\
y                   & 0.948$^{+0.611}$            & 0.399$^{+0.703}$   & 1.407$^{+0.570}$            \\
\hline
caustic crossings   & 2.430$^{+0.809}$   & 0.276$^{+0.462}$   & 1.929$^{+0.710}$   \\
\hline
\end{tabular}
\end{table}

\section{Discussion}
In the next few years LSST and Euclid will transform the field of quasar microlensing by discovering orders of magnitude of new lensed quasars.
\cite{OM10} estimate that a total of $\sim$8000 such systems will be found, $\sim$3000 of which will be bright enough to be detected in a single LSST visit, thus enabling the direct extraction of light curves through its 10-year-long monitoring campaign.
Considering the estimates of $\sim$600 quads, $\sim$5 ``naked'' cusps and $\sim$2400 doubles \citep{OM10}, this amounts to a total of $\sim$7500 lensed quasar images potentially monitored by the LSST.
Therefore, even though high magnification microlensing events in lensed quasars are a rare phenomenon (see Table \ref{tab:nevents}), the sheer number of images being observed will make them fairly common.

In order to quantify the yields of LSST observing strategies, in this work we focus on three systems covering a broad range of characteristics. 
The main factor for selecting specifically these three systems is their location in the sky (see Fig. \ref{fig:obsvel}), which directly affects their relative velocity with respect to the CMB and the number and time intervals between LSST observations.
A secondary yet still important factor, is the relative redshifts between lens and source, which determine the size of the Einstein radius and hence the microlensing scale length.
Finally, the amount of smooth matter fraction at the location of each quasar image, which directly affects the number of caustics on a magnification map, is very different for each system.
Based on the \citet{OM10} prediction of the lensed quasar redshift distribution having a peak at $z_l = 0.65 \pm 0.27$ and $z_s = 2.29 \pm 0.86$, and the (commonly) high smooth matter mass fraction expected \citep{pooley2012}, \he is the most representative system in our sample.

Unsurprisingly, the average magnitude of the transverse velocities shown in Fig. \ref{fig:angvelsim} is consistent with the values computed by \cite{mosquera_kochanek2011}, however, we show here that the direction of these velocities (fully accounted for in this work) can make a difference.
Indeed, the $\boldsymbol{\upsilon}_o$ component of the effective velocity given in equation (\ref{eq:veff_explicit}) has a fixed transverse direction with respect to the system location in the sky.
Given that magnification maps are not homogeneous due to the fixed direction of the shear (towards the center of mass of the lens in the absence of external shear, which usually coincides with its center of light), sets of randomly drawn light curve trajectories cannot be treated as completely equivalent and may significantly diverge in their sampled properties.
In Fig. \ref{fig:angvelsim} we see this effect on the velocity distributions for \he, where the large projected velocity of the CMB (a $\sim47$ per cent of the total) is sufficient to break the uniform distribution in terms of the velocity magnitude.
On the other hand, for \rxj, which is located in a low $\boldsymbol{\upsilon}_{\rm CMB}$ region, the random direction velocity components dominate.
Interestingly, even though \q does not lie in a very low $\boldsymbol{\upsilon}_{\rm CMB}$ region (a non-negligible $\sim18$ per cent of the total) the large distance between the lens and the source leads to a sufficiently high projected $\boldsymbol{\upsilon}_{\rm CMB}$ component.
Hence, despite the velocity distribution appearing also skewed, this is not as prominent as it is for \he.
Given that \he is the most representative of the three systems, we conclude that this effect needs to be taken into consideration when examining microlensing light curves.

High magnification microlensing events, albeit rare, are very powerful in constraining the structure of accretion discs, especially when they correspond to caustic crossings \citep[e.g.][]{anguita08}.
\cite{mosquera_kochanek2011} present the expected Einstein crossing times - the amount of time the source takes to cross one Einstein radius $R_{\rm E}$ in the source plane - for all lensed quasars known at the time of publication.
Estimated at an average of $\sim$20 years, this time scale is usually considered as the average time between high magnification events/caustic crossings. However, as shown in Table \ref{tab:nevents}, the number of high magnification events and even their relation with the number of caustic crossings cannot be directly estimated from the average transverse velocity due to: i) the specific strong lens parameters ($\kappa, \gamma, s$) that have a direct impact on the density of caustics within the magnification pattern; ii) the angular dependence of the velocity, making caustic crossing encounter probabilities not uniformly distributed due to the alignment between $\phi_\gamma\text{[deg]}$ and $\boldsymbol{\upsilon}_{\mathrm{CMB}}$ (if the system is in a high $|\boldsymbol{\upsilon}_{\mathrm{CMB}}|$ region) ; and iii) the accretion disc size and thermal profile, which make an important difference between systems and observing bands, i.e., smaller sources are more likely to produce magnification events above our magnitude cut than larger sources (as can be seen in Table \ref{tab:nevents}). Note that since we adopted a scale-invariant accretion disc model (a self-similar power law), results obtained in different observing bands are due only to changes in disc size (e.g. measuring microlensing properties in a redder band is equivalent to selecting a larger $r_0$ in a bluer one).

These differences are made evident in the cases of \he and \rxj.
Due to their macromodel parameters, they have the lowest (high $s$; minimum) and highest (low $s$; highest magnification saddle) caustic density, respectively.
In the case of \he, this naturally results in the lowest number of high magnification events and caustic crossings.
The low caustic density implies that the events are dominated by single caustic or cusp interactions and thus have a low sensitivity to observing band (source size).
Another consequence of the low caustic density is that a sizeable fraction of the events are not due to caustic crossings but due to cusp passings.
As such, even when the average Einstein crossing time (from the average transverse velocity and the Einstein radius) is $\approx$20 years, the average caustic crossing time is $\approx$36 years.
This also explains the lower probability density of high magnification events in the u filter, where the source is smaller, compared to g and r, where the source becomes increasingly larger and more affected by nearby cusps (see middle panel of Fig. \ref{fig:eventdays}).
On the other hand, \rxj, with a very high caustic density, has the largest number of caustic crossings despite having a significantly longer Einstein crossing time than \q (which has a higher average transverse velocity).
However, the number of high magnification events is lower than in \q and, in fact, lower than the number of expected caustic crossings.
This is explained by the larger projected source size (in R$_E$): very close caustics are not ``resolved'' in the light curves even in the u-band (smallest source), thus, multiple caustic crossings can result in single high magnification events.
In contrast to \he, the inferred average Einstein time for \rxj is $\approx$16 years whereas the average caustic crossing time is significantly lower than that: $\approx$4 years.
The number of high magnification events, in this case, is significantly more sensitive to observing band (size), and longer than in \he in spite of the larger average transverse velocity.

In summary, the properties of high magnification events (e.g. expected number, duration, and nature), besides directly depending on the average transverse velocity, have the following first order dependencies with the input parameters:
\begin{itemize}[leftmargin=*]
    \item Number of caustic crossings \textrightarrow  Macro model ($\kappa, \gamma, s$).
    \item Number of high magnification events \textrightarrow  Macro model ($\kappa, \gamma, s$) and accretion disc ($r_0$, $\nu$).
    \item Event duration \textrightarrow  Accretion disc ($r_0$, $\nu$).
    \item Event slope \textrightarrow  Accretion disc size and nature of the event (caustic crossing, cusp passing).
\end{itemize}

In our ``average'' system \he, the probability of a high magnification event with $\Delta$mag$>$1.0 mag in the LSST bands (in only one of the multiple images) is of $>$0.047 per year.
This amounts to at least 0.047$\times$7500$=\sim$350 $\Delta$mag$>$1.0 mag microlensing events per year for all the lensed quasar images observed by LSST.

In Fig. \ref{fig:opsimplots} we show the performance of LSST regarding the detection of such high magnification events for several observing strategies.
Since quasar microlensing events are generally long, we can see that with the uniform (non-rolling) cadences, most of the events are detected for all systems.
However, when we move towards rolling cadences, we see mixed results.
In particular, those rolling cadences that completely neglect portions of the sky for an entire observing season in all filters (e.g. mothra\_2049 and alt\_sched\_rolling) are those that perform worst.
However, unless a very extreme/exotic observing strategy is executed by LSST, the majority of the events will be detected.
We do note, however, that mere detection of isolated high magnification events does not by itself guarantee accurate constraints on accretion disc sizes and thermal profiles.
Significantly more stringent constraints are obtained by accurately sampling the magnification event \citep[e.g.][]{kochanek2004,eigenbrod08,anguita08,morgan10}.
Considering that the typical event duration in our representative system \he is between 1 and 4 months (depending on observing band, see Fig. \ref{fig:eventdays}), most uniform cadence experiments would not achieve more than a few observations per event (especially in the u band, that has the shortest event duration and the longest median separation of the observations).
As such, it will become of paramount importance to predict high magnification events on a sparse baseline sample to trigger higher cadence follow up of the events.

Although our results are appropriate to broadly estimate the number and properties of high magnification events potentially detected by LSST for comparing different observing strategies in ``typical'' systems, there are several caveats in the analysis that should be noted.
Firstly, we used identical stellar velocity dispersions for the lensing galaxies and (intrinsic frame) dispersions for the peculiar velocities of the lens and the source, even though they are located at different redshifts.
These, however, have negligible contributions compared to the relative redshift differences.
Secondly, we have assumed an identical temperature profile for the three systems examined.
Thirdly, our experiment focused only on the pure microlensing variability of one image per system, ignoring intrinsic variability coupled with time delays. However, in close pairs of images or highly symmetric quadruple systems such an effect can safely be ignored (this is not the case for the remaining systems, whose numbers are not negligible). Finally, we have not analyzed in depth the considerations that arise from distinguishing microlensing signal variability in minima and saddle-point images \citep[see][]{schechter02}.

To allow for reproducibility, further testing, and quick adoption of our results, we give access to the tools produced as part of this work: i) an automated online tool that can send computation requests to a supercomputer, returning the results to the user via a download link; ii) decompression and plotting programs; iii) the backend LSST code; and iv) the backend simulations code that can run either on GPU or CPU.

\section*{Acknowledgements}
FN and TA acknowledge support from Proyecto FONDECYT N: 1190335. GV is supported through an NWO-VICI grant (project number 639.043.308). The authors would like to acknowledge the contribution of H. Witt, who was kind enough to share his caustic finder code.

\bibliographystyle{mnras}
\bibliography{LSST} 

\bsp	
\label{lastpage}
\end{document}